\documentclass[twocolumn,showpacs,preprintnumbers,amsmath,amssymb,bibnotes]{revtex4}

\usepackage{graphicx}
\usepackage{dcolumn}
\usepackage{bm}
\newcommand{\bl}{{[\![}}
\newcommand{\sint}{\,{\textstyle\int}}
\newcommand{\br}{{]\!]}}
\newcommand{\be}{\begin{equation}}
\newcommand{\ee}{\end{equation}}
\newcommand{\comment}[1]{}

\begin{document}

\preprint{LMU--ASC 49/06}

\title{Amplitude for $\bm{N}$-Gluon Superstring Scattering}%

\author{Stephan Stieberger}
\affiliation{
Arnold--Sommerfeld--Center for Theoretical Physics,\\
Ludwig--Maximilians--Universit\"at M\"unchen,\\
Theresienstra\ss e 37, 80333 M\"unchen, Germany}

\author{Tomasz R. Taylor}
\affiliation{
Department of Physics, Northeastern University\\
Boston, MA 02115, United States of America
}%

\begin{abstract}
We consider scattering processes involving $N$ gluonic massless states of open superstrings with 
certain Regge slope $\alpha'$. At the semi-classical level, the string world-sheet sweeps a disk 
and $N$ gluons are created or annihilated at the boundary. We present exact expressions for the 
corresponding amplitudes, valid to all orders in $\alpha'$, for the so-called maximally helicity 
violating configurations, with $N{=}4,\; 5$ and
$N{=}6$. We also obtain the leading ${\cal O}(\alpha'^{\,2})$ string corrections to the zero-slope 
$N$-gluon Yang-Mills  amplitudes.
\end{abstract}

\pacs{11.25.Wx, 11.30.Pb, 12.38.Bx}
\maketitle

In gauge theories, the scattering of gauge bosons reveals, in the most direct way,  the structure of fundamental 
interactions.
In particular, in Quantum Chromodynamics (QCD), the scattering of gluons yields the abundance of hadronic jets. 
Experimental studies of multi-jet production at hadron colliders provide some of the most convincing tests of QCD.
Techniques for calculating scattering amplitudes had been
steadily developed for the last thirty years, with an important progress achieved in the 1980's, during the research 
and development stage of the Superconducting Super Collider. In a couple of years, the Large Hadron Collider (LHC) 
will start running, and while we are very well prepared for
testing QCD and the rest of the standard model of particle physics, there are
certainly more scenarios for physics beyond the standard model than envisaged twenty years ago.

Although many models have been proposed, there is no clear prediction for the energy scale of new physics. In open (type I) superstring theory, this scale is determined by the so-called Regge slope $\alpha'$ of mass dimension ${-}2$. Massless gauge bosons are separated by a mass gap of $1/\sqrt{\alpha'}$  from the massive string modes.
Traditionally, the Regge slope
and the respective string mass scale had been tied to the Planck mass, however more recently, some serious 
consideration has been given to models with much lower string mass scale \cite{Lykken:1996fj,Antoniadis:1998ig},
possibly even within the reach of LHC.
The full string amplitudes depend on $\alpha'$, resulting in large
corrections to Yang-Mills amplitudes once some kinematic invariants characterizing energy scales involved in the 
scattering process become comparable to $1/\sqrt{\alpha'}$. Such deviations from Yang-Mills theory are due
to virtual heavy string modes and in principle, they can be observed well below the string threshold as a signal of new physics.

In Yang-Mills theory, there exists a subclass of amplitudes that are described, at the tree-level, by
a simple analytic formula valid for arbitrary number $N$ of external gauge bosons. Hereafter, for abbreviation, 
generic gauge bosons will be called gluons and the corresponding gauge group the color group. We consider amplitudes 
with all momenta directed {\it inward}. Assume that two gluons, with the momenta
$k_1$ and $k_2$, in the color states described 
by the matrices $\lambda^1$ and $\lambda^2$, respectively, in the fundamental representation of the gauge group, carry negative 
helicities while the rest of gluons, with the momenta and color charges $(k_3,\lambda^3),\dots,(k_N,\lambda^N)$, respectively,
carry positive helicities.
Then the scattering amplitude for such a ``maximally helicity violating'' (MHV) configuration is given by
\cite{Parke:1986gb,Berends:1987me}
\begin{equation}
\mathfrak{M}_{Y\!M}^{(N)}=ig^{N-2}\ {\rm Tr}(\lambda^1\cdots\lambda^N)\
\frac{\langle 12\rangle^4}{\langle 12\rangle\langle 23\rangle\cdots\langle N1\rangle}
\label{pt}\end{equation}
where $g$ is the gauge coupling constant and $\langle ij\rangle$ are the spinor products
\cite{Mangano:1990by,Dixon:1996wi} of the momentum space wavefunctions
describing
free left-handed fermions with the momenta equal to  $k_i$ and $k_j$, respectively \cite{precise}.

In this Letter, we focus on MHV amplitudes describing multi-gluon scattering in
open superstring theory, with the expectation that Eq.(\ref{pt}) has a simple generalization to all orders in $\alpha'$.
All tree diagrams including exchanges of both massless particles as well as heavy string states, are generated
from just one string diagram: a disk world-sheet with $N$ string vertices creating (or annihilating) gluons at the 
boundary.
This Letter is intended as a summary of results; details of the computations are given elsewhere
\cite{Stieberger:2006te}.

In order to write down the amplitudes in a concise way, it is convenient to introduce the following
notation for the kinematic invariants characterizing $N$-particle scattering: \begin{eqnarray}\label{in}
\bl i\br_n&=&\alpha'\, (k_i+k_{i+1}+\dots+k_{i+n})^2\\
&&\hskip -1.4cm\epsilon(i,j,m,n)\;=\, \alpha'^{\, 2}\,\epsilon_{\alpha\beta\mu\nu}\,
k_i^{\alpha}k_j^{\beta}k_m^{\mu}k_n^{\nu}\,  \label{pseudo}\end{eqnarray}
with the cyclic identification
$i+N\equiv i$. Here, $\epsilon_{\alpha\beta\mu\nu}$ is the four-dimensional Levi-Civita symbol.
All gluons are on-shell, {\it i.e}.\ $k_i^2=0$.
The factors of $\alpha'$ render the above invariants dimensionless.

The amplitude for four-gluon scattering has been known for a long time \cite{GS,Schwarz,VectorT}.
All string effects are summarized in one Euler function (Veneziano amplitude) as
the formfactor of Yang-Mills amplitude, so that \begin{equation}\label{m4}
\mathfrak{M}^{(4)}=\frac{\Gamma(1+s_1)\ \Gamma(1+s_2)}{ \Gamma(1+s_1+s_2)}\ \mathfrak{M}_{Y\!M}^{(4)}\ ,
\end{equation}
where $s_1\equiv\bl 1\br_1$, $s_2\equiv\bl 2\br_1$. Note that in
the low-energy limit, the leading correction to Yang-Mills amplitude appears at order
${\cal O}(\alpha'^{\,2})$. It can be interpreted
as the effect of a contact interaction term $\alpha'^{\, 2}{\rm Tr}\, F^4$, with the Lorentz indices
of four gauge field strengths $F_{\mu\nu}$ contracted by the well-known $t_{(8)}$ tensor.

The five-gluon MHV amplitude can be extracted from the recent calculations
\cite{Barreiro:2005hv,Medina:2002nk}. In this scattering process, five invariants
are neccesary to specify the kinematics. They can be chosen as
$s_i\equiv\bl i\br_1,~i=1,\dots,5$, {\it i.e}.\ as the  cyclic  orbit of
$\bl 1\br_1$ obtained by the action of 
the cyclic group 
$\mathbb{Z}_5$ generated by $i\to i+1$ ({\it mod} 5).
The computations involve integrations over two vertex positions:
\be
\int_0^1\!\!\! dx\!\!\int_0^1\!\!\!dy\ x^{s_2}y^{s_5}(1-x)^{s_3}(1-y)^{s_4}(1-xy)^{s_1-s_3-s_4}R(x,y)\, ,\nonumber
\ee
with various rational functions $R(x,y)$. Although such integrals can be expressed in terms of
certain hypergeometric functions,  it is more convenient to label them by $R(x,y)$.
We use the following shorthand notation: let $\sint R(x,y)$ denote the above integral
evaluated for a given function $R(x,y)$.
One finds 
\cite{Stieberger:2006te} that the amplitude can be expressed in terms of just two functions: 
\begin{equation}\label{m5}
\mathfrak{M}^{(5)}=
[\ V^{(5)}(s_i)-2i\ P^{(5)}(s_i)\ \epsilon(1,2,3,4)\ ]\ \mathfrak{M}_{Y\!M}^{(5)}\ ,\end{equation}
where
\begin{eqnarray}\label{pv5}
P^{(5)}(s_i)&=& \sint (1-xy)^{-1}\ ,\nonumber \\
V^{(5)}(s_i)&=& s_2s_5\sint (xy)^{-1}
\\ \;+\;&&\hskip -7mm \frac{1}{2}\ (\ s_2s_3+s_4s_5-s_1s_2-s_3s_4-s_1s_5\ )\ P^{(5)}(s_i)\, .\nonumber
\end{eqnarray}
The above functions, as well as the pseudoscalar $\epsilon(1,2,3,4)$, are cyclic \cite{cyc},
thus the full factor in front of the Yang-Mills amplitude $\mathfrak{M}_{Y\!M}^{(5)}$ is cyclic invariant. The origin of the 
pseudoscalar part can be explained by the presence of the ${\rm Tr}\, F^4$ effective interaction term. 
The low-energy behavior of the amplitude
(\ref{m5}) is determined, up to the order ${\cal O}(\alpha'^3)$, by the following expansions:
\begin{eqnarray}\label{v5ex}
P^{(5)}(s_i)&=& \frac{\pi^2}{6}-\zeta(3)\ \{s_1\}+\dots\ ,\\ \nonumber
V^{(5)}(s_i)&=&  1-\frac{\pi^2}{12}\ \{s_1s_2\}\\ &&+~ \nonumber\frac{\zeta(3)}{2}\  
(\ \{s_1^2s_2\}+\{s_1s_2^2\}+\{s_1s_3s_5\}\ )+\dots
\end{eqnarray}
where the curly brackets enclosing kinematic invariants imply the summation over all distinct 
elements of the respective cyclic orbits \cite{cexpl}.

The connection to the four-gluon amplitude (\ref{m4}) can be established by
considering the so-called soft limit, $k_j\to 0$. Then the pseudoscalar part of the factor
disappears due to the momentum conservation while the function
$V^{(5)}(s_i)\to\Gamma(1+s_1) \Gamma(1+s_2) [\Gamma(1+s_1+s_2)]^{-1}$, reproducing the Veneziano formfactor of (\ref{m4}).

Next, we proceed to the case of six gluons.
Although in four dimensions, only eight kinematic invariants are necessary to parametrize six-particle
scattering, it is convenient to consider
an extended nine-element set that is more natural from the point of view of $\mathbb{Z}_6$ cyclic 
symmetry:
$s_i\equiv\bl i\br_1,~i=1,\dots,6$ and $t_j\equiv\bl j\br_2,~j=1,2,3$,
{\it i.e}.\ the $\mathbb{Z}_6$ orbits of $\bl 1\br_1$ and $\bl 1\br_2$.
These variables are subject to a fifth-order polynomial constraint \cite{asribekov} that can be ignored
here because it does not play any direct role in our
considerations. The computations involve integrations over three vertex positions:
\begin{widetext}\vskip-0.5cm
\begin{equation}
\int_0^1\!\!\! dx\!\!\int_0^1\!\!\!dy\!\!\int_0^1\!\!\!dz\  x^{s_2}y^{t_2}z^{s_6}
(1-x)^{s_3}(1-y)^{s_4}(1-z)^{s_5}(1-xy)^{t_3-s_3-s_4}(1-yz)^{t_1-s_4-s_5}
(1-xyz)^{s_1+s_4-t_1-t_3}R(x,y,z).\nonumber\end{equation}
Here again, we label integrals by the rational functions $R(x,y,z)$, with  $\sint R(x,y,z)$
denoting the above integral evaluated for a given  $R(x,y,z)$.
Using as a starting point the results of \cite{Oprisa:2005wu}, we obtain
\begin{eqnarray}\nonumber
\label{m6}
\mathfrak{M}^{(6)}&=&
\left[\ V^{(6)}(s_i,t_i)-2i\, P^{(6)}_1(s_i,t_i)\ \epsilon(2,3,4,5)
-2i\, P^{(6)}_2(s_i,t_i)\ \epsilon(1,3,4,5)\right.\\ &&\hskip 6mm\left. -2i\, 
P^{(6)}_3(s_i,t_i)\ \epsilon(1,2,4,5)-2i\, P^{(6)}_4(s_i,t_i)\ \epsilon(1,2,3,5)
-2i\, P^{(6)}_5(s_i,t_i)\ \epsilon(1,2,3,4)\ \right]\ \mathfrak{M}_{Y\!M}^{(6)}\ ,
\end{eqnarray}
\vskip-0.4cm
\begin{eqnarray}\nonumber
{\rm \hskip-1.3cm with\ the\ functions:\ \ \ \ }   
P^{(6)}_1&=& s_1\sint [(1-xy)\ (1-yz)\ (1-xyz)]^{-1}+(s_2+s_5-t_1-t_2)\sint [(1-xy)\ (1-xyz)]^{-1}\\[.5mm]
&&\nonumber \hskip 5mm +~(s_6+s_5-s_1-t_2)\sint yz\ [(1-yz)\ (1-xyz)]^{-1}\ ,\\
P^{(6)}_2&=& s_2\sint [x\ (1-xy)\ (1-yz)]^{-1}+(s_3+s_6-t_2-t_3)\sint y\ [(1-xy)\ (1-yz)]^{-1}\ ,\nonumber\\
P^{(6)}_3&=& s_3\sint [(1-x)\ (1-yz)]^{-1}+(s_1+s_4-t_1-t_3)\sint yz\ [(1-yz)\ (1-xyz)]^{-1}\ ,\\
P^{(6)}_4&=& s_4\sint [(1-y)\ (1-xyz)]^{-1}+(s_2+s_3-s_4-t_2)\sint [(1-xyz)]^{-1}\ ,\nonumber\\
P^{(6)}_5&=& s_5\sint [(1-z)\ (1-xy)]^{-1}+(s_3+s_4-s_5-t_3)\sint [(1-xy)(1-xyz)]^{-1}\ ,\nonumber
\end{eqnarray}
\begin{eqnarray}
V^{(6)}&=&s_2s_5t_2\sint [xy\ (1-z)]^{-1}+
\frac{1}{2}\ (s_2 s_3 - s_3 s_4 + s_3 s_6 + s_4 t_2 - s_2 t_3 - t_2 t_3)\ P^{(6)}_1\nonumber\\ &&
\hskip -3mm +\,\frac{1}{2}\ 
(-s_2 s_3 + s_1 s_4 - s_4 s_5 - s_3 s_6 + s_3 t_1 - s_4 t_2 
+ s_2 t_3 + s_5 t_3 - t_1 t_3 + t_2t_3)\ P^{(6)}_2\\ &&\hskip -3mm +\,\frac{1}{2}\ 
(s_2 s_3 - s_1 s_4 + s_2 s_5 + s_3 s_6 + s_5 s_6 - s_3 t_1 - s_6 t_1  - s_2 t_3 - 
  s_5 t_3+ t_1t_2  + t_1t_3 - t_2t_3)\ P^{(6)}_3\nonumber\\ &&\hskip -3mm +\,\frac{1}{2}\ 
(-s_2 s_3 + s_1 s_4 - s_2 s_5 - s_1 s_6 + s_3t_1 + s_6t_1 + s_1t_2+ s_2 t_3 - 
t_1t_2  - t_1t_3)\ P^{(6)}_4\nonumber\\ &&\hskip -3mm +\,\frac{1}{2}\ 
(-s_1 s_2 + s_2 s_3 + s_2 s_5 - s_3 t_1 - s_1 t_2 + t_1t_2)\ P^{(6)}_5-s_5s_3\ P^{(6)}_2+
s_5(s_3-t_2)\ P^{(6)}_3
\nonumber
\end{eqnarray}
Although the above functions seem to be complicated, they have very simple transformation 
properties under cyclic permutations. In particular, $V^{(6)}(s_i,t_i)$ is cyclic invariant
while the functions $P^{(6)}(s_i,t_i)$ transform among themselves in such a way that the imaginary part
of the prefactor in Eq.(\ref{m6}) is also invariant \cite{Stieberger:2006te}. As a result, similarly to the case
of four and five gluons, the full string formfactor of the MHV six-gluon amplitude (\ref{m6})
is cyclic invariant. It also has the correct soft limits when any momentum goes to zero:
\begin{eqnarray}\label{factorize}
V^{(6)}(s_i,t_i)&\xrightarrow[k_j= 0]{}&V^{(5)}(s_i)\ \ \ ,\ \ \ 
\sum_{l=1}^5 (-1)^{l+1}\ P^{(6)}_l(s_i,t_i)\xrightarrow[k_6= 0]{}P^{(5)}(s_i)\\ \nonumber
P^{(6)}_l(s_i,t_i)&\xrightarrow[k_l= 0]{}&P^{(5)}(s_i)\ \ \ ,\ \ \ l=1,\ldots,5\ ,
\end{eqnarray}
factorizing into the infrared pole times the five-gluon amplitude (\ref{m5}).
The low-energy 
behavior of the amplitude  (\ref{m6}) is determined, up to the order ${\cal O}(\alpha'^3)$, by the following expansions:
\begin{eqnarray}\label{lowlimits}\nonumber
V^{(6)}(s_i,t_i)&=& 1-\frac{\pi^2}{12}\ (\ \{s_1s_2\}-\{s_1s_4\}+\{t_1t_2\}\ )+
\frac{\zeta(3)}{2}\ \left(\ \{s_1s_2^2\}+\{s_1^2s_2\}-\{s_1^2s_4\}+\{s_1s_2t_1\}\right.\\ \nonumber
&&\left. 
-\{s_1s_4t_1\}-\{s_2s_5t_1\}-3 \{s_1s_4t_2\}+\{s_1t_1t_3\}+\{t_1t_2^2\}+\{t_1^2t_2\}+3\, t_1t_2t_3\ 
\right)+\ldots\ ,\\ \nonumber
P^{(6)}_1(s_i,t_i)&=&
\frac{\pi^2}{6}+\zeta(3)\ (s_1 + 2 s_2 - s_3 - s_4 + 2 s_5 + s_6 - 3 t_1 - 3 t_2 - t_3)+
\ldots\ ,\\ 
 P^{(6)}_2(s_i,t_i)&=&
\frac{\pi^2}{6}+\zeta(3)\  (2 s_2 + 2 s_3 - s_4 - s_5 + s_6 - t_1 - 3 t_2 - 2 t_3)+
\ldots\ ,\\ \nonumber
P^{(6)}_3(s_i,t_i)&=&
\frac{\pi^2}{6}+\zeta(3)\ (2 s_3 + s_4 - s_5 - s_6 - t_1 - t_2 - 2 t_3)
+\ldots\ ,\ \\ \nonumber
P^{(6)}_4(s_i,t_i)&=&
\frac{\pi^2}{6}+\zeta(3)\ (-s_1 + s_3 + s_4 - s_6 - t_1 - t_2 - t_3)
+\ldots,\\ \nonumber
P^{(6)}_5(s_i,t_i)&=&
\frac{\pi^2}{6}+\zeta(3)\ (-s_1 - s_2 + s_3 + 2 s_4 - t_1 - t_2 - 2 t_3)+\ldots\ .
\end{eqnarray}
\end{widetext}

It is clear from the discussion of $N=4,5$, and especially of $N=6$, that the complexity of amplitudes increases with $N$. The integrals become more complicated and the number of 
independent functions grows. The functions emerging in the step from $N{-}1$ to $N$ have low-energy 
expansions beginning with $\zeta(N{-}3)$ and they proliferate at
order ${\cal O}(\alpha'^{N-3})$. However, if one is interested in a fixed order in $\alpha'$, then only a 
limited number of functions is relevant. In particular,
at ${\cal O}(\alpha'^{\,2})$,
it is sufficient to expand
$V^{(N)}$ up to quadratic order and set all $P^{(N)}\approx\pi^2/6$. At this order, the scattering process has a particularly simple effective field 
theory description
in terms of Feynman diagrams centered at one single ${\rm Tr}F^4$ interaction vertex
and trees of gluons spreading from there, with gluons multiplying through Yang-Mills
interactions {\it i.e}.\ via Altarelli-Parisi decays \cite{Stieberger:2006te}. Below, we  write down a simple expression for the 
corresponding amplitude.

An $N$-particle scattering process can be parametrized in terms of $N(N-3)/2$
kinematic invariants which can be chosen as the cyclic orbits of $\textstyle
\bl 1\br_k,~k=1,\dots,E(\frac{N}{2}-1)$, where $E$ denotes the integer part.
Recall that the $\mathbb{Z}_N$ group of cyclic permutations is generated by the shift of indices labeling gluons from 
$i\to i+1$ $({\it mod}\, N)$.
Note that for $N$ odd, the last orbit contains $N$ elements, while for $N$ even their number
is reduced  by the momentum conservation to $N/2$. As in the case of $N=6$, we can ignore the four-dimensional 
constraints \cite{asribekov} that reduce the number of independent invariants to $3N-10$.

Up to the leading ${\cal O}(\alpha'^{\,2})$ correction, the $N$-gluon MHV superstring amplitude has the form
\be\label{mn}
\mathfrak{M}^{(N)}=\big[\ 1-\frac{\pi^2}{12}\, Q^{(N)}\ \big]\
\mathfrak{M}_{Y\!M}^{(N)},\ee
where $Q^{(N)}$ are Lorentz-invariant, homogenous of degree four, functions of the momenta.
They are uniquely determined by two requirements: cyclic symmetry and soft limit,  $Q^{(N)}\to Q^{(N{-}1)}$ as $k_N\to 0$.
As a result of iteration to arbitrary $N$, we obtain
\begin{eqnarray}\nonumber
Q^{(N)}&=& \sum_{k=1}^{E(\frac{N}{2}-1)}\{\,\bl 1\br_k\bl 2\br_k\}-\!\!
\sum_{k=3}^{E(\frac{N}{2}-1)}\{\,\bl 1\br_k\bl 2\br_{k-2}\}\\
&& +~C^{(N)}+\, 4i \hskip - 4mm \sum_{k<l<m<n<N}\!\!\!\epsilon(k,l,m,n)\, , \label{mhvn}\\[2mm]
&&\hskip -11mm C^{(N)} = \left\{ \begin{array}{ll}
        -\{\,\bl 1\br_{\frac{N}{2}-2} \bl \frac{N}{2}+1\br_{\frac{N}{2}-2}\}  
& \mbox{$N>4$, 
even,}\\[2mm]
         -\{\,\bl 1\br_{\frac{N-5}{2}} \bl \frac{N+1}{2}\br_{\frac{N-3}{2}}\} & \mbox{$N>5$, odd}.
\end{array}\nonumber \right.
\end{eqnarray}
Here again, the curly brackets enclosing kinematic invariants imply the summation over all distinct 
elements of their cyclic orbits.
As a check, it is easy to verify  that the result (\ref{mhvn}) agrees with Eqs.(\ref{m4}), (\ref{m5}) and (\ref{m6}).  
We believe that after classifying
the integrals over an arbitrary number of gluon vertex positions,
a similar iterative procedure can be used to determine the full
MHV superstring amplitudes for all $N$ \cite{prep}.

The results presented here hold for any superstring compactification from
ten to four dimensions, with ${\cal N}=1$ or higher supersymmetry, or with supersymmetry broken by $D$-brane 
configurations because at the disk level, supersymmetry breaking does not communicate to multi-gluon interactions.
The ${\cal O}(\alpha'^{\,2})$ corrections of Eq.(\ref{mhvn}) apply also to any theory
like supersymmetric technicolor in which the
dimension eight operator ${\rm Tr}F^4$ is induced at the compositess scale
$\Lambda\approx 1/\sqrt{\alpha'}$. 
Actually, they bear a 
striking resemblance to the one-loop all positive helicity
amplitudes of QCD \cite{Bern:1993qk}. Hence
it would be interesting to investigate a possible type I--heterotic duality \cite{Witten:1995ex,Polchinski:1995df} relation of our results to Ref.\cite{Bern:1993qk} and to the recent 
computations of all one-loop MHV amplitudes in QCD \cite{Berger:2006vq}.
It would be also very interesting
to understand if there is any room in the twistor space formulation of string theory 
\cite{Witten:2003nn} (see also \cite{Cachazo:2005ga,Dixon:2005cf})    
that would allow accomodating superstring corrections to Yang-Mills scattering amplitudes.

Hopefully, LHC will reach beyond the standard model and the signals of new physics will rise above the QCD background.
But even if no spectacular effect, like a direct production of massive string modes, is discovered, some threshold effects may be observed in multi-jet production, due to the presence of contact interactions induced by virtual particles too heavy to be produced on-shell. If this is the case, then our results could be important for LHC data analysis.
\vskip 4mm
\begin{acknowledgments}
\vskip -0.4cm
We are grateful to Lance Dixon for very useful correspondence.
In addition, we wish to thank K.S.\ Narain and Dan Oprisa for discussions.
This work  is supported in part by the European Commission under Project MRTN-CT-2004-005104.
The research of T.R.T.\ is supported in part by the U.S.
National Science Foundation Grants PHY-0242834 and PHY-0600304.
T.R.T.\ is grateful to Dieter L\"{u}st,
to Arnold Sommerfeld Center for Theoretical Physics at Ludwig Maximilians
University, and to Max-Planck-Institut in Munich, for their kind hospitality. 
He would like to also thank the Galileo Galilei Institute for Theoretical Physics for 
hospitality and INFN for partial support during completion of this work.
Any opinions, findings, and conclusions or
recommendations expressed in this material are those of the authors and do not necessarily
reflect the views of the National Science Foundation.
\end{acknowledgments}

\bibliography{gluons}

\begin{thebibliography}{25}
\bibitem[{cyc()}]{cyc}
\bibinfo{note}{In discussing cyclic transformations, we always take into
  account the $\mathbb{Z}_N$-invariant momentum conservation constraint
  $\sum_{i=1}^{i=N}k_i=0$}.
\end{thebibliography}

\end{document}